\documentclass{iaus}
\usepackage{graphicx}

\title[Precessing AGN Jets, Bubbles and Cooling Flows] 
{Precessing AGN Jets, Bubbles and Cooling Flows}

\author[D. Falceta-Gon\c calves et al.]   
{D. Falceta-Gon\c calves$^{1,2}$, A. Caproni$^2$, Z. Abraham$^3$, E. M. de Gouveia Dal Pino$^3$ \and D. M. Teixeira$^3$}

\affiliation{$^1$Escola de Artes, Ci\^encias e Humanidades, Universidade de S\~ao Paulo - Rua Arlindo Bettio 1000, CEP: 03828-000, S\~ao Paulo, Brazil \\[\affilskip]
$^2$N\' ucleo de Astrof\'
isica Te\' orica, Universidade Cruzeiro do Sul - Rua Galv\~ ao Bueno
868, CEP 01506-000, S\~ao Paulo, Brazil \\[\affilskip]
$^3$Instituto de Astronomia, Geof\'\i sica e Ci\^encias Atmosf\'ericas, 
Universidade de S\~ao Paulo, Rua do Mat\~ao 1226, CEP 05508-900,
S\~ao Paulo, Brazil}

\pubyear{2010}
\volume{275}  
\pagerange{x--x}
\jname{Jets at all Scales}
\editors{A.C. Editor, B.D. Editor \& C.E. Editor, eds.}
\begin{document}

\maketitle

\begin{abstract}

Several galaxy clusters are known to present multiple and misaligned pairs of cavities seen 
in X-rays, as well as twisted kiloparsec-scale jets at radio wavelengths. 
It suggests that the AGN precessing jets play a role in the formation of the misaligned bubbles. 
Also, X-ray spectra reveal that typically these systems are also able to supress cooling flows, predicted 
theoretically. The absence of cooling flows in galaxy clusters has been a mistery for many years since 
numerical simulations and analytical studies suggest that AGN jets are highly energetic, but are unable to 
redistribute it at all directions. We performed 3D hydrodynamical simulations of the interaction between a 
precessing AGN jet and the warm intracluster medium plasma, which dynamics is coupled to 
a NFW dark matter gravitational potential. Radiative cooling has been taken into account and the cooling flow 
problem was studied. We found that precession is responsible for multiple pairs of bubbles, as observed. The misaligned 
bubbles rise up to scales of tens of kiloparsecs, where the thermal energy released by the jets are redistributed. 
After $\sim 150$ Myrs, the temperature of the gas within the cavities is kept of order of $\sim 10^7$ K, while the denser 
plasma of the intracluster medium at the central regions reaches $T \sim 10^5$ K. The existence of multiple bubbles, at 
diferent directions, result in an integrated temperature along the line of sight much larger than the 
simulations of non-precessing jets. This result is in agreement with the observations. The simulations 
reveal that the cooling flows cessed $\sim 50 - 70$ Myr after the AGN jets are started.

\end{abstract}

\firstsection 

\vspace*{0.5 cm}

Several "S"-shaped structures are observed at radio wavelengths emerging from AGNs. These structures are well known to be produced by precessing jets. Some AGNs are located at the central region of dense galaxy clusters, as NGC 1275, at the Perseus Galaxy Cluster and Hydra A.

The low density but high temperature intracluster medium (ICM) is observable in X-rays. The most intriguing features of the X-rays emission maps are the presence of cavities and voids (Sanders \& Fabian 2007). In some cases, detached and misaligned pairs of cavities are observed. Theoretical models predict that AGN jets – at certain circumstances – can inflate bubbles that buoyantly rise up to the outer parts of the cluster core. However, it is not clear how misaligned cavities can be formed. Theoretically, the ICM free-free emission is responsible for a fast cooling of the gas at the cluster core, resulting in a "cooling flow". The cooling flows, however, have not yet been detected. AGN feedback would therefore help to prevent the cooling flows. 

The main goal of this work was to simulate precessing jets of AGNs and obtain density and temperature distributions that explain both the observed x-rays and synchrotron emission maps, as well as the suppression of cooling flows. In order to do so, we used a Godunov-scheme hydrodynamic code with a fixed grid, with $512^3$ cartesian cells (Burkhart et al. 2009, Falceta-Goncalves, Lazarian \& Houde 2010). The radiative cooling is set implicitly by the use of a cooling function $\Lambda (T)$ (Le\~ao et al. 2009). The computational domain is of $L_{box}=100$ kpc in each direction. The ICM was initially set as in hydrostatic equilibrium with a Navarro-Frenk-White dark matter distribution. The code solves the mass and momentum hydrodynamic equations including the gravitational potential of the dark matter. The jet is started at $t=0$ with $n=5  \times 10^{-3}$ cm$^{-3}$, $T=10^9$K, $v_{jet}=10^4$km s$^{-1}$.

The jet precesses around the axis of the total angular momentum, considering both the accretion disk and black hole angular momenta. We run a number of simulations with different initial precession angles in the range $\varphi_0 = 0 - 60^{\rm o}$, and its period was set as $T_{prec} = 1.5 \times 10^7$yr.

The simulations were calculated up to $t=150$ Myr, which represents $3 T_{prec}$. As in several previous similar works, we found that for precessing angles $\varphi_0 < 50^{\rm o}$ the jet carves the ICM creating elongated cavities. As the system evolves the jet reaches farther distances of the AGN. The cooling at the core of the ICM is responsible for a cooling flow that does not cease. However, for $\varphi_0 > 50^{\rm o}$, the jet inflated cavity becomes buoyantly unstable and rises up of the cluster core. Due to the precession, when it happens the jet is pointing towards a different direction and starts to inflate a different pair of bubbles. The result is the formation of several misaligned pairs of bubbles. The larger filling factor of the hot and low density cavities, compared to non-precessing jets, is the main cause of the cooling flow stagnation. In Fig. 1 we present the mass flux of the ICM gas for the model with $\varphi_0 = 60^{\rm o}$. It is clear that right after $t=1 T_{prec}$ the formation of the cavities drastically reduce the cooling flow (see Falceta-Goncalves et al. 2010a,b for more details).

\begin{figure}[]
\begin{center}
 \includegraphics[width=3.5in]{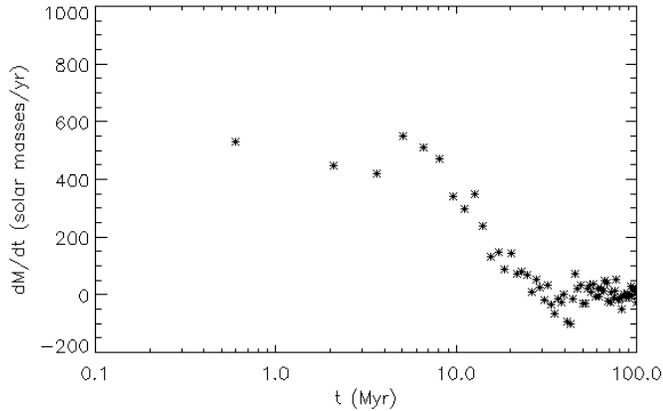} 
 \caption{Mass flow during the simulation. Positive values correspond to inward fluxes.}
   \label{fig1}
\end{center}
\end{figure}

\end{document}